\documentclass[aps,superscriptaddress,amsmath,amssymb,twocolumn,amsfonts]{revtex4-1}
\usepackage{graphicx}
\usepackage{hyperref}
\usepackage{cleveref}
\usepackage{color}
\usepackage{float}
\usepackage{soul}
\usepackage{ulem}
\usepackage{braket}
\usepackage{dsfont}
\usepackage{physics}
\usepackage{tensor,upgreek,float}
\usepackage{parskip}

\begin{document}

\title{Thermodynamics of vison crystals in an anisotropic quantum spin liquid} 
\author{Ritwika Majumder}
\affiliation{School of Physical Sciences, National Institute of Science Education and Research, A CI of Homi Bhabha National Institute, Jatni 752050, India}

\author{Onur Erten}
\affiliation{Department of Physics, Arizona State University, Tempe, Arizona 85281, USA}

\author{Anamitra Mukherjee}
\affiliation{School of Physical Sciences, National Institute of Science Education and Research, A CI of Homi Bhabha National Institute, Jatni 752050, India}


\begin{abstract}
Using unbiased Monte Carlo simulations and variational analysis, we present the ground state and finite temperature phase diagrams of an exactly solvable spin-orbital model with Kitaev-type interactions on a square lattice. We show that an array of new gapped and gapless vison crystals -- characterized by the periodic arrangement of $\mathbb{Z}_2$ flux excitations -- can be stabilized as a function of external magnetic field and exchange anisotropy. In particular, we discover a variety of `quarter phases' wherein new sixteen-site periodic patterns emerge, with only a quarter of the fluxes adopting 0-flux configurations. In contrast, the rest remain in $\pi$-flux configurations. Vison crystals break translational symmetry and undergo finite temperature phase transitions. We investigate the finite temperature properties of these phases and report the corresponding critical and crossover temperatures. Our results reveal an array of novel phases in exactly solvable extensions of the Kitaev model, wherein local and topological orders can coexist.

\end{abstract}

\maketitle

\section{Introduction}
Unlike conventional magnets where the magnetic moments align to create long-range magnetic order, in quantum spin liquids (QSLs), the spins do not form long-range ordered patterns even when absolute zero temperature is approached\cite{Broholm_Science2020, Balents_Nature2010, Savary_RepProgPhys2016,moessner_moore_2021}. On the contrary, QSLs exhibit long-range entanglement, fractionalization of low-energy excitations, and emergent gauge fields due to their topological properties, which have become the defining characteristics of QSLs\cite{Zhou_RMP2017, Knolle_AnnRevCondMatPhys2019, Wen_RMP2017}. In the pursuit of the realization of QSLs, the Kitaev model on honeycomb lattice\cite{kitaev-2006} plays a crucial role as it is the first exactly solvable model featuring fractionalized abelian and non-abelian excitations. These excitations are promising candidates for quantum error correction protocols, vital for the realization of quantum computers\cite{Cui_Quantum2020}. Despite the experimental progress in identifying materials with strong Kitaev interactions such as iridates \cite{Kitagawa_Nat2018}, RuCl$_3$ \cite{Takagi_NatRevPhys2019} and CrI$_3$ \cite{Lee_PRL2020}, a definitive confirmation of a Kitaev QSL state remains elusive.

The exact solution of the Kitaev model has been critical to uncover the detailed nature of $\mathbb{Z}_2$ QSLs. This solution involves non-interacting Majorana fermions that are coupled to gauge invariant fluxes threading through the plaquettes constructed from the bond-dependent gauge fields. In specific lattices such as honeycomb and square lattice, the ground state flux configuration follows the Lieb's theorem\cite{lieb-1994}. The thermodynamic properties of the Kitaev model can be studied within Monte Carlo approaches in varying dimensions\cite{kee-1d-mc, burnell-kitaev-mc, motome-3d-mc}. 

\begin{figure}[t]
\centering{
\includegraphics[width=7.0cm, height=4.75cm, clip=true]{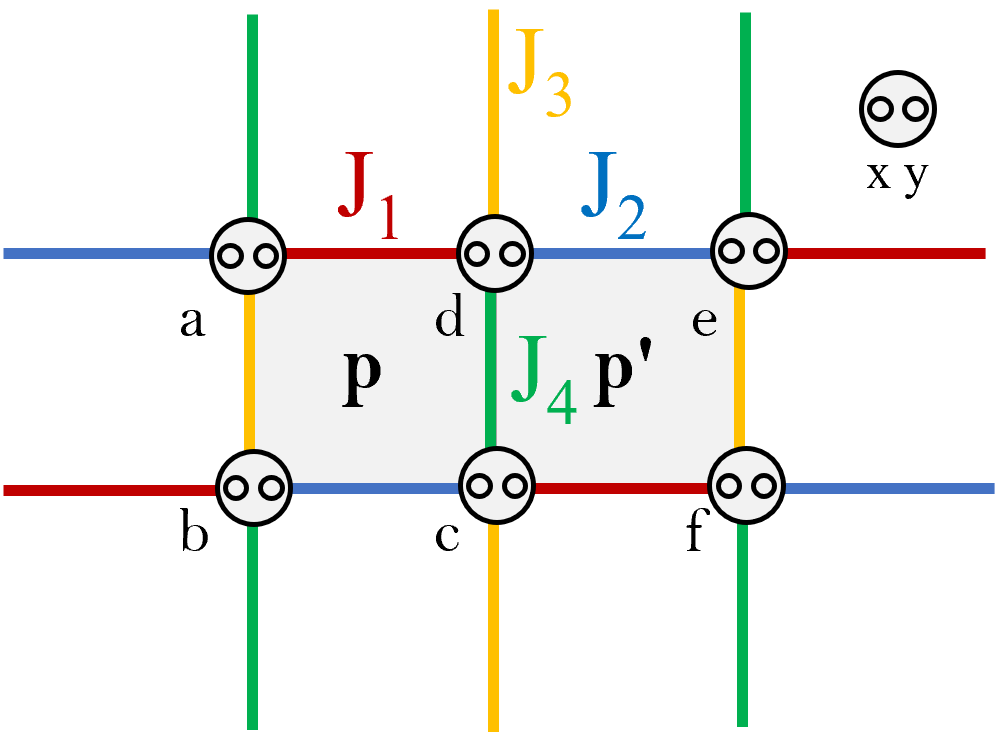}} 
\caption{Schematic of the model with Kitaev-type interactions on a square lattice. We identify two inequivalent plaquettes, $p$ and $p^\prime$. The Majorana fermion representation of the spin and orbital degrees of freedom results in two flavors of free Majorana fermions that are coupled to static $\mathbb{Z}_2$ fluxes.}
\vspace{-0.0cm}
\label{f-1}
\end{figure}
A key ingredient in the exact solution of the Kitaev model is the anticommutation properties of Pauli matrices, $\{\sigma_i, \sigma_j\} =\delta_{ij}$. Since there are only three Pauli matrices, Kitaev model can only be applied to tricoordinated lattices such as honeycomb, hyperhoneycomb and hyperoctagon lattices. Additionally, this limits the perturbations that can be added to the Kitaev model, which preserves its integrability. For instance, the exact solution is lost in the presence of an external magnetic field. However, the exact solution can be maintained in the presence of certain extended Kitaev interactions. Ref.~\citenum {batista-2d-mc} showed that in this case, Lieb's theorem does not apply, and the ground state can shift from zero-flux configuration to a variety of vison crystals. Vison crystals are formed by the periodic arrangement of the flux excitations. Since they break translational symmetry, they exhibit local Ginzburg-Landau-Wilson type order parameters. In that sense, they show a resemblance to Abrikosov vortex lattices or skyrmion crystals. However, the former are the periodic condensation of the topological defects of order parameters, whereas visons crystals are the the periodic arrangement of gapped flux excitations of $\mathbb{Z}_2$ gauge theories, in the absence of a local order parameter.

An alternative way to tune the ground state properties of Kitaev-type QSLs is to consider the $\Gamma$-matrix generalizations\cite{Wu_PRB2009} of the Kitaev model. For instance, for a four-dimensional representation of the $\Gamma$ matrices, there are five anti-commuting operators, $\Gamma^\alpha$, with $\{\Gamma^\alpha, \Gamma^\beta\} = \delta_{ij}$. This allows to construct exactly-solvable models with coordination number, $z$, up to $z=5$ \cite{Chern_PRB2010} or models that exhibits a higher symmetry\cite{yao-lee-flux-lattice, Nakai_PRB2012, Vijayvargia_PRR2023, onur-moire}. Additionally, the remaining $\Gamma^\alpha$ operators can be included as local terms without destroying the integribility\cite{vojta-flux-lattice-1}. While variational analysis and Monte Carlo simulations have been recently applied to uncover vison crystals in a variety of models with Kitaev-type interactions\cite{vojta-flux-lattice-0,onur-quasicrystal, Akram_PRB2023}, finite temperature properties of vison crystals models have received limited attention \cite{batista-2d-mc}.

Hence, we study the ground state and finite temperature properties of an exactly solvable model with Kitaev-type interactions on a square lattice via unbiased Monte Carlo simulations and variational analysis. We discover a variety of new vison crystals as a function of anisotropic exchange and external magnetic field. In particular, we uncover different quarter (Q) phases where new 16-site periodicity emerge. We report the temperature dependence of the static flux structure factors, crossover and transition temperatures, and the construct the finite T phase diagram as a function of external magnetic field and exchange anisotropy.

The paper is organized as follows. We discuss the model and the Monte Carlo approach in Sec. II, along with the observable quantities computed. In Sec. III A, we present the vison crystal phases for different kinds of exchange anisotropies and magnetic fields at zero temperature. We also present the properties of gapped and gapless spin-orbital liquid phases as a function of magnetic field and anisotropy. In Sec. III B, we discuss the temperature-field phase diagrams for various anisotropies by studying the evolution of the gauge-invariant static flux structure factor. In Sec. IV, we discuss the observables for identifying the signatures of flux crystals and conclude with a summary.

\begin{figure*}[t]
    \centering{
    \includegraphics[width=1\linewidth]{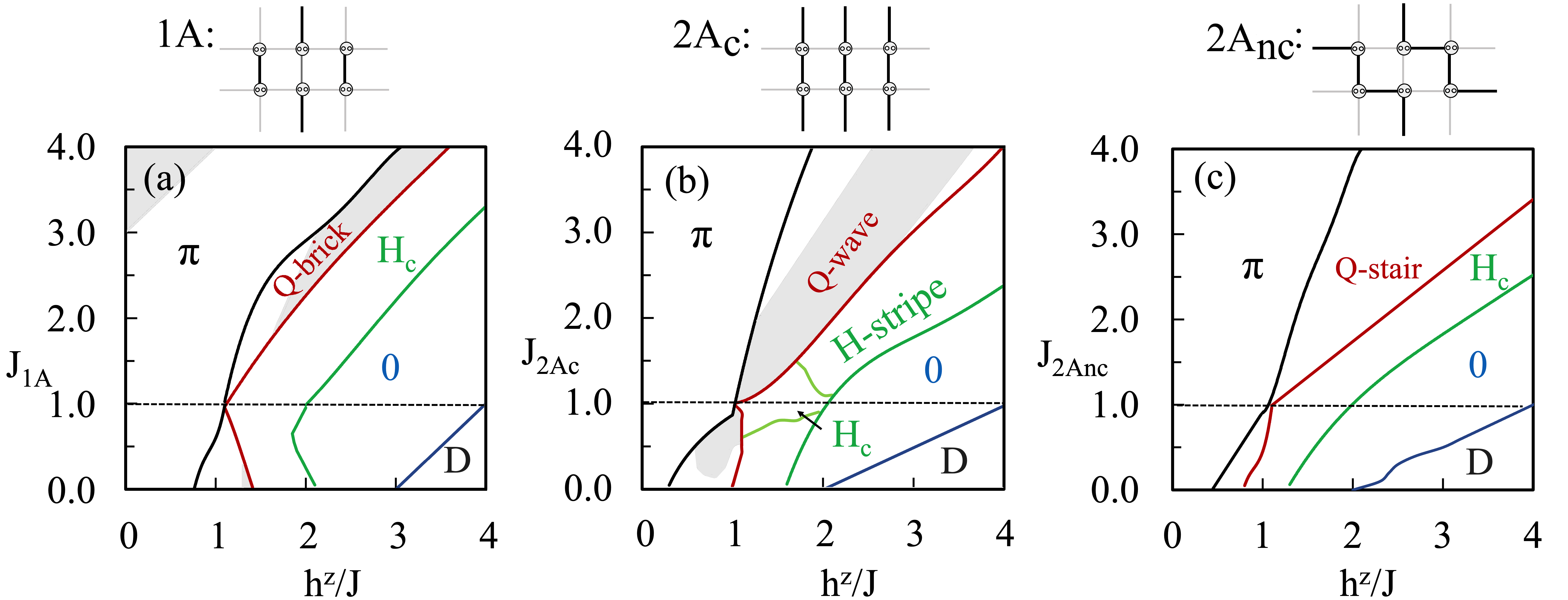}}
    \caption{Ground state phase diagram for different types of anisotropic exchange and external magnetic field: (a) to (c) show the vison crystal phases for one-bond ($1A$), two-bond colinear $(2A_{c})$ and two-bond non-colinear $(2A_{nc})$ exchange anisotropies, respectively. The schematic of the anisotropies is shown at the top of each phase diagram. The anisotropy grows with the deviation of the ratio on the vertical axis from the isotropic value of 1 in all three cases. The phases at the isotropic point are denoted by $\pi$, H$_c$ (half-flux-checkerboard), and $0$, with uniform flux values -1, staggered flux arrangement, and zero or flux-free, respectively, in all three panels. D denotes the degenerate flux phase which is unstable to confinement. With deviation from the isotropic limit, new quarter-flux phases are stabilized between the $\pi$ and the H$_c$ phases in all panels. These quarter-flux vison crystals have distinct lattice periodicity and are denoted as Q-brick, Q-stair, and Q-wave in the panels. We also find a new half-flux phase (H-wave) with periodicity distinct from the half-flux checkerboard phase H$_c$ for two-bond colinear $(2A_{c})$. In panels (a) and (b), the shaded regions depict the gapped phases, and all other phases are gapless.}
    \label{f-2}
\end{figure*}
\section{Microscopic Model \& method}
We consider an exactly solvable model with Kitaev-type interactions in the presence of an external magnetic field, originally introduced by Ref.~\citenum{Nakai_PRB2012}. The Hamiltonian comprises of spin ($\sigma$) and orbital ($\tau$) degrees of freedom on a square lattice and nearest neighbor bond-dependent spin-orbital exchanges $J_{\gamma}$, $\gamma\in\{1,2,3,4\}$ as depicted in Fig.\ref{f-1}. 

\begin{figure*}[t]
    \centering{
    \includegraphics[width=1\linewidth]{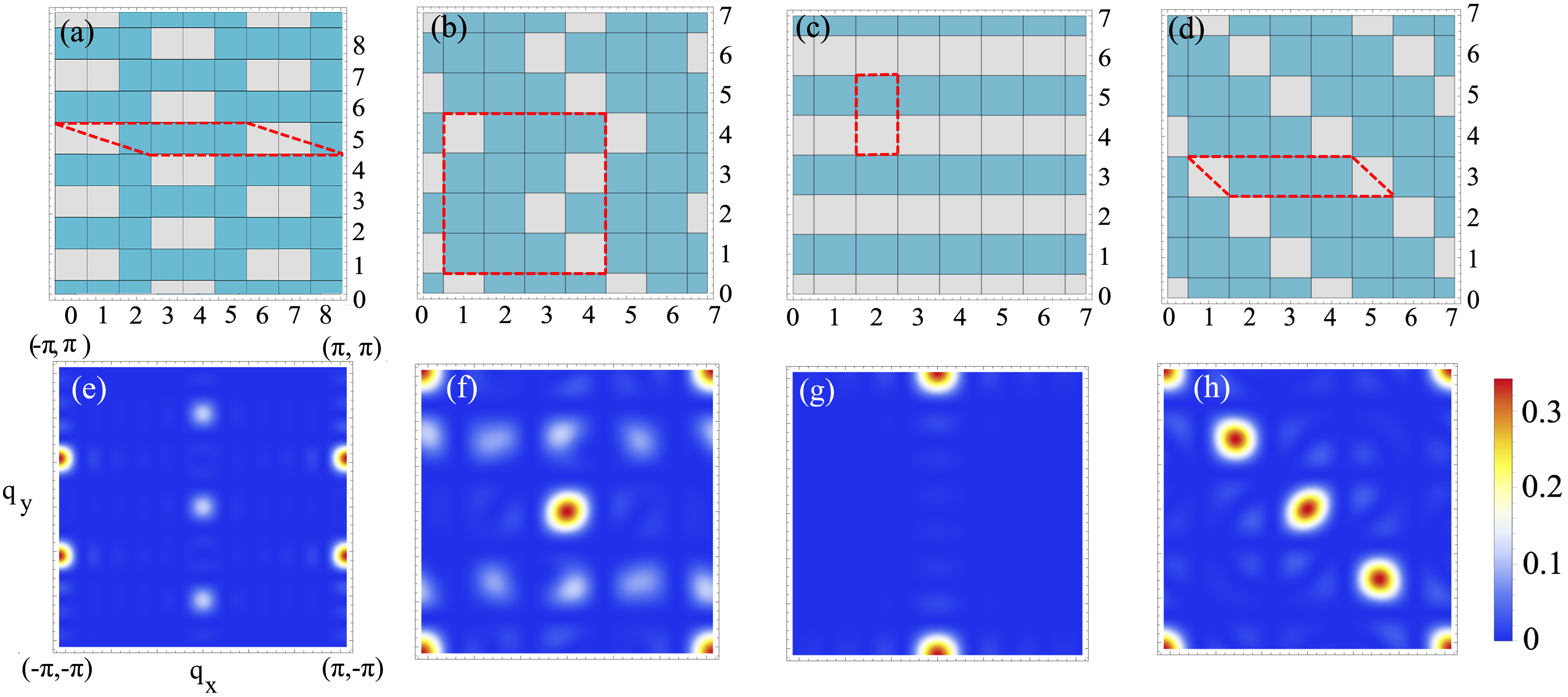}}
    \caption{\label{fig: wide} Snapshots of the ground state vison crystals and static flux structure factors: (a)  and (b) show the flux values at the plaquette centers of the square lattice for the quarter phases: Q-brick and Q-stairs. Blue and gray plaquettes indicate a flux value -1 and +1 respectively. The plaquette centres are labelled in the panels. Panels (e) and (f) show the corresponding static flux structure factors. Panels (c) and (d) show the new half-flux and quarter-flux Q-wave phases, respectively. Panels (g) and (h)  show the corresponding static flux structure factors S$(\mathbf{q})$. The dashed red boundaries in the top panel depict the unit cells for the respective cases.}
    \label{f-3}
\end{figure*} 

\begin{eqnarray}
    \label{e1}
    H&=&-\sum_{\langle ij\rangle_{\gamma}}J_{\gamma} (\sigma^{x}_i\sigma^{x}_j+\sigma^{y}_i\sigma^{y}_j)\otimes\tau_i^{\gamma}\tau_j^{\gamma} \nonumber \\
    &&-h^z\sum_{i}\sigma_i^z\otimes \mathds{1}.
\end{eqnarray} 
$\sigma$ and $\tau$ are spin and orbital Pauli matrices, resulting in a four-dimensional local Hilbert space. Note that the Hamiltonian exhibits U(1) symmetry in the spin sector. The exact solution of Eq.~\ref{e1} relies on the anticommution relations of the $4\times 4$ $\Gamma$-matrices satisfying Clifford algebra and the above model is realized with the identification  $\Gamma^{a}=-\sigma^y\otimes\tau^{a}$ for $(a=1,2,3)$, $\Gamma^4=\sigma^x\otimes\mathds{1}$, $\Gamma^5=-\sigma^z\otimes\mathds{1}$ \cite{Wu_PRB2009, vojta-flux-lattice-1}. 
We label two inequivalent plaquette operators $W_{p}=\sigma^z_{d}\sigma^z_{ c}\otimes\tau^x_{ a}\tau^y_{b}\tau^x_{c}\tau^y_{d}$ and $W_{ p^\prime}=\sigma^z_{d}\sigma^z_{ c}\otimes\tau^x_{c}\tau^y_{d}\tau^x_{e}\tau^y_{ f}$, each with $\pm$ 1 eigenvalues. $W_{p/p^\prime}$ commute with the Hamiltonian, including the external magnetic field term. The Hilbert space is divided into sectors of conserved fluxes. 
Next, we employ a Majorana fermion representation, $\Gamma^{\alpha}_i= ib_i^{\alpha}c_i$, and denote $b_i^5\rightarrow c_i^x$ and $c_i\rightarrow c_i^y$ for convenience and obtain
\begin{eqnarray}
      \label{e2}
    \mathcal{H}=\sum_{\langle ij\rangle_{\gamma}}J_{\gamma} u^{\gamma}_{ij}(ic^x_ic^x_j+ic^y_ic^y_j)+h^z\sum_i ic^x_ic^y_i  
\end{eqnarray}
where $u_{ij}^{\gamma}=-ib^{\gamma}_ib^{\gamma}_j$. Similar to the Kitaev model, the bond operators commute with the Hamiltonian $(\mathcal{H})$ in extended Hilbert space, and plaquette operators, $W_{p/p^\prime}$ can be expressed as a product of the $u_{ij}^{\gamma}$ around the plaquette. The Hamiltonian is invariant under a $\mathbb{Z}_2$ gauge transformation $\{c_i^x, c_i^y\}\rightarrow -\{c_i^x, c_i^y\}$; $u_{ij}\rightarrow -u_{ij}$. The Hilbert space of the Majorana fermions is overcomplete and subject to the constraint $D_i=i c_i^x c_i^y b_i^1 b_i^2 b_i^3 b_i^4 =1$. The wave function can be projected to the physical Hilbert space by $P=\prod_i(1+D_i)/2$, as $P|\Psi\rangle = |\Psi\rangle_{\rm Phys}$. The Majorana fermion representation of the Hamiltonian in Eq.~\ref{e2} entails two flavors of Majorana fermions hopping on a square lattice that is coupled to static (bond-dependent) $\mathbb{Z}_2$ gauge field, along with an external magnetic field that couples the two flavors. 

According to Lieb's theorem, the ground state of Eq.~\ref{e2} for $h^z=0$ lies in the $\pi$-flux sector\cite{lieb-1994}. However, for finite $h^z$, Lieb's theorem does not apply, and we perform Monte Carlo simulations together with variational analysis to estimate the ground state and finite temperature phase diagram. The Monte Carlo scheme consists of an exact diagonalization (ED) of the Majorana fermions in the background of the static $\mathbb{Z}_2$ gauge fields. The free energy of the fermions is used to thermally anneal the $\mathbb{Z}_2$ fluxes using the standard classical Monte Carlo (MC) with a Metropolis algorithm. The ED+MC method is a common technique to study models that consists of classical degrees of freedom that are coupled to free fermions such as metallic magnets\cite{Erten_PRL2011}. Previous work on vison crystals\cite{batista-2d-mc} and finite temperature properties of Kitaev model also utilize the ED+MC method\cite{Nasu_PRB2015, burnell-kitaev-mc, motome-3d-mc}. In our simulations, we consider $24\times12$ lattices and start at a high temperature with random flux configurations. Next, we gradually cool down the system and perform 4000 Monte Carlo sweeps for each temperature. We estimate the ground state flux configuration at $T = 10^{-3}J$.
We track the ground state and its thermal evolution with $M=\Sigma_i\langle \sigma_i^z\rangle$, with $i$ running over all lattice sites, and static flux structure factor defined as
\begin{equation}\label{e3}
    S(\textbf{q})=\langle \frac{1}{N^2}\sum_{p,p^\prime}W_p W_{p^\prime} e^{-i\textbf{q}\cdot(\textbf{r}_p-\textbf{r}_{p^\prime})}\rangle
\end{equation}
where $p, p^\prime$ denotes the position of the plaquette centers. These observables are averaged over 300 configurations for every temperature, leaving ten system sweeps to avoid self-correlations. These observables allow us to track the thermal evolution of fermions as well as excitations of the flux ground states. The energy difference among the vison crystals can be quite small and vanishes at the first-order phase transitions. Therefore, Monte Carlo simulations suffer from phase separation. To overcome this issue, we also construct variational states motivated by the vison crystals obtained in the Monte Carlo simulations. We use these variational states to obtain the ground state phase diagram.

To characterize the flux-phases, we also define a \textit{flux-polarization per unit cell} as the ratio of the number of plaquettes with positive flux threading it to the number of plaquettes in the unit cell for a given flux phase. In the discussion below, all half flux-phases are prefixed with `H' and all quarter-flux phases by `Q.'

\section{Results}
\subsection{Ground state phase diagram}
We begin our analysis with the ground state phase diagram as a function of magnetic field ($h^z$) and exchange anisotropy. We consider three types of anisotropic bond configurations: a single bond anisotropy ($1A$) where one of the $J_\gamma$ is different from the rest and two types of double bond anisotropies $2A_c$ and $2A_{nc}$, with two colinear $\{J_1=J_2\} \neq \{J_3=J_4 \}$ and two perpendicular bonds $\{J_1=J_3\} \neq \{J_2=J_4 \}$ are anisotropic respectively. A schematic of these anisotropic exchanges is shown in Fig.~\ref{f-2}. We define an anisotropy parameter $J_{1A}\equiv J_a/J$, with $J_1=J_2=J_3=J$ and $J_4=J_a$ for $1A$. Similarly  $J_{2A_c}=J_a/J$ with $J_1=J_2=J_a$ and $J_3=J_4=J$ and  $J_{2A_{nc}}=J_a/J$ with $J_1=J_3=J_a$ and $J_2=J_4=J$. The isotropic cases correspond to $J_{1A}$, $J_{2A_c}$ and $J_{2A_{nc}}$ equal to 1. Fig.~\ref{f-2} (a) to (c) show the ground state phase diagram as a function of magnetic field for the one-bond (1A), two-bond colinear (2A$_c$) and two-bond non-colinear (2A$_{nc}$) anisotropies respectively.

Our Monte Carlo simulations are in agreement with the previously reported variational calculations in the isotropic limit\cite{vojta-flux-lattice-1}. For $h^z=0$, the ground state is in the $\pi$-flux sector as dictated by Lieb's theorem, where the flux through all the plaquettes takes the uniform value of $W_p=-1$. We find that $\pi$-flux extends up to $h^z/J\lesssim 1.1$. For $1.1 <h^z/J\lesssim 2$  a checkerboard half-flux phase (H$_c$) is stabilized followed by $0$-flux phase for $2<h^z/J\leq 4$. For $h^z> 4$, all vison configurations become degenerate, indicating gapless vison excitations for which the topological order is unstable to confinement\cite{vojta-flux-lattice-1}. We find that all of the phases remain gapless in the isotropic limit up to $h^z/J\leq4$. 

The Monte-Carlo calculations also reproduce the expected $S(\mathbf{q}=(0,0))$ structure factor peaks for the $\pi$ and $0$ flux vison crystals and the $\mathbf{q}=(\pi,\pi)$ peak for the checkerboard phase in the isotropic limit. In addition, we also reproduce the first-order transitions between these phases as a function of the magnetic field. Additionally, the magnetization for these phases grows monotonically with $h^{z}$, with discontinuous jumps at the first order boundaries, eventually saturating to the maximum value of 1 at large $h^z>4J$ in agreement with literature. As per the flux-polarization nomenclature discussed in the previous section, the checkerboard phase is denoted by H$_c$, while the standard nomenclature of $\pi$ and zero flux-phases are retained.

Next, we discuss the effects of anisotropic exchange on the vison crystal phase diagrams. The $\pi$ flux phases become more stable as anisotropy is increased beyond 1. As a result, phase boundaries in all cases shift to a higher magnetic field as shown in Fig.~\ref{f-2}. In all three cases, we discover new quarter flux phases (Q-brick, Q-wave, and Q-stair) situated between the $\pi$ and the half-flux vison crystals where the periodicity of Q phases differ in each case. The schematics of these phases and the corresponding static flux structure factors are shown in Fig.~\ref{f-3}. In all of these phases, a quarter of plaquettes in the unit cell exhibit 0-flux, whereas the rest have $\pi$-flux configuration. For small anisotropy, the windows of these phases widen with increasing magnitude of anisotropy. The field dependence of these quarter-flux vison crystals, however, is markedly different, with anisotropy greater than 1, requiring a larger field to stabilize these phases for all three cases. For anisotropy less than 1, for  1A  in Fig.~\ref{f-2}(a), we observe that the Q-brick phase spreads uniformly about the field value 1.1, the transition field value between $\pi$ and H$_c$ in the isotropic limit. For the Q-wave and Q-stair phases, the stability window is pushed progressively to smaller fields with an anisotropy ratio approaching zero. In addition, we find a new half-flux phase H-stripe for 2A$_c$ beyond the critical anisotropy window as indicated in panel (b). 

The unit cell for the vison crystals Q-brick, Q-wave, H-stripe, and Q-stair are marked in red in the upper panels in Fig.~\ref{f-3} in panels (a) through (d), respectively. The structure factor peaks characterizing the Q-brick, Q-wave and Q-stairs are $\{(0,0), (0,2\pi/3), (\pi,\pi/3)\}$, $\{(0,0), (\pi,\pi), (\pi/2,\pi/2), (0,\pi/2),(\pi,\pi/2)\}$ and $\{(0,0), (\pi/2,-\pi/2), (\pi,-\pi)\}$ respectively, shown in panels  (e), (f) and (h) respectively. The relative ratio of the peak magnitudes is $(1/6, 1/6,1/3)$  for the Q-brick phase, $(1/4,1/4,1/8,1/8)$ for the Q-wave phase, and equal $1/4$ weight for all four peaks for Q-stair phase. The H-stripe phase in panel (g) has sharp peaks at $\mathbf{q}=(0,-\pi)/(0/\pi)$ in the structure factor with equal amplitudes.

\begin{figure*}[t]
    \centering{
    \includegraphics[width=1.0\linewidth]{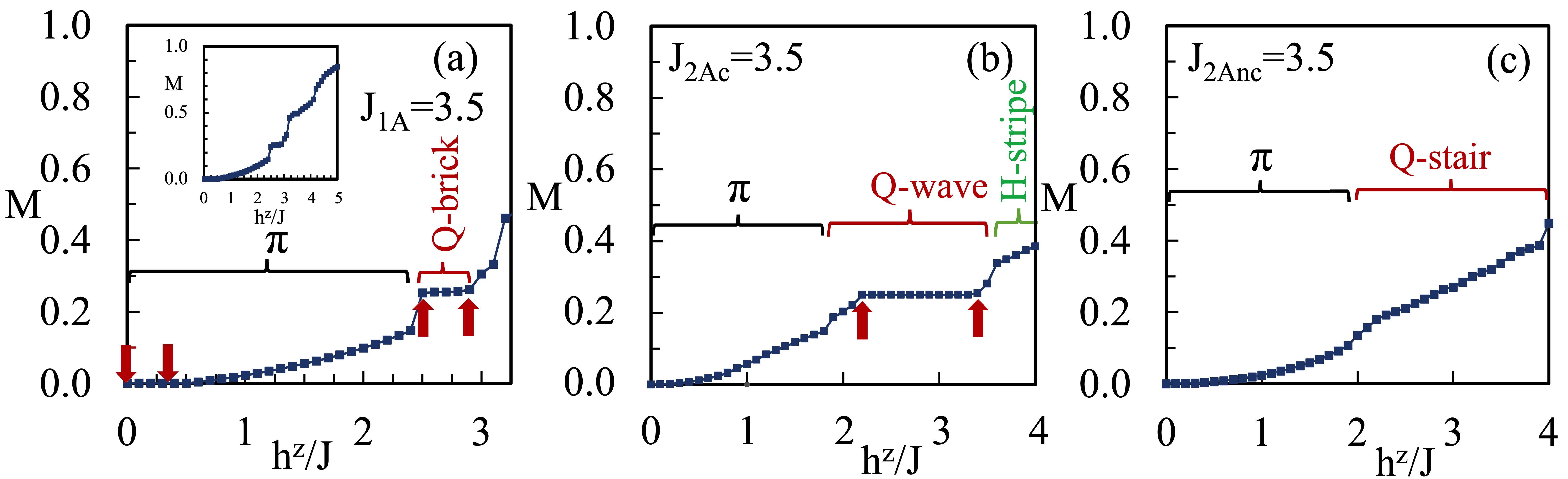}}
    \caption{\label{fig: wide}{Evolution of magnetization with magnetic field in the ground state:} (a - c) show the magnetization ($M$) for 1A, 2A$_c$ and 2A$_{nc}$ respectively for fixed magnitude of anisotropy 3.5. The gapped regimes of the vison crystals are indicated by the region between pairs of arrows, while the full range of the vison crystal phases is marked. Inset in (a) shows the data in the main panel over a wider magnetic field range and is discussed in the text. }
    \label{f-4}
\end{figure*}

Using the analytical expressions for  Majorana fermion excitation spectrum, we determine if the vison crystals are gapped or gapless. We find that the majority of them are gapless whereas the gapped visons crystals are shown in the shaded regions of Fig.~\ref{f-2}. The presence of a gap in the Majorana fermion spectrum suppresses the magnetic susceptibility. As a result, the magnetization as a function of $h^z$ show plateaus for the gapped phases as shown in Fig.~\ref{f-4}. On the contrary, magnetization increases linearly with $h^z$ for the gapless phases. Even though Fig.~\ref{f-4} only shows a fixed anisotropy parameter, 3.5, the conclusions drawn from hold in general.  The M($h^z$) is expected to grow with increasing magnetic field, which tends to polarize the spins. However, in gapped vison crystal phases, the gap in the Majorana spectrum prevents the increase of magnetization, resulting in a distinct magnetization plateau of magnitude of 0.25 for the quarter and 0.5 for the gapped vison crystals. In Fig.~\ref{f-4} (a), we see that the magnetization exhibits plateaus for  $0<h^z<0.5$ with M=0 and for $2.5<h^z<2.9$ with M=0.25. These are coincident with the gapped $\pi$ and Q-brick vison crystals in Fig.~\ref{f-2} (a). The inset of  Fig.~\ref{f-4} (a) also shows the evolution of the magnetization over a wide range of magnetic field values. In Fig.~\ref{f-4} (b), the Q-wave vison crystal has both gapless and gapped regimes for anisotropy 3.5. In  Fig.~\ref{f-4} (b), we see that the magnetization increases monotonically with the magnetic field in the gapless Q-wave vison crystal regime but exhibits a distinct plateau with M=0.25 in the gapped regime. In  Fig.~\ref{f-4} (c), we find that M increases in a strictly monotonic fashion with the field as there are no gapped phases in Fig.~\ref{f-2} (c). 

\begin{figure*}
    \centering
    \includegraphics[width=1.0\linewidth]{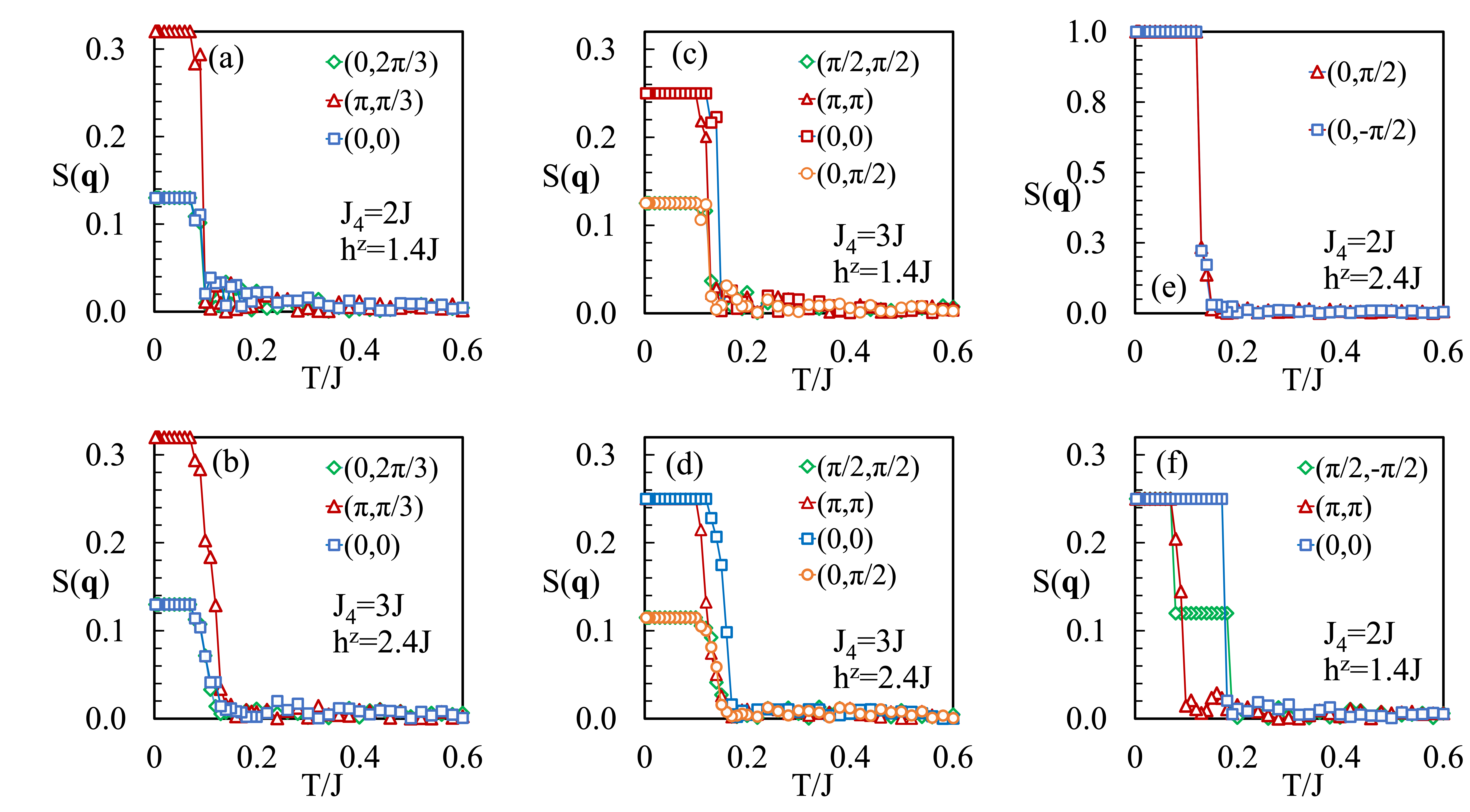}
   
    \caption{{Temperature evolution of structure factors for various gapped and gapless phases:} (a) and  (b) show the temperature evolution of the structure factor for the Q-brick phase for a gapless (a) and gapped (b) vison crystal. (c) and (d) show the structure factor evolution for gapped (c) and gapless (d) Q-wave phase vison crystal. (e)  and (f) show the temperature evolution of $S(\mathbf{q})$ for the gapless H-stripe and Q-stair phases respectively. }
    \label{f-5}
\end{figure*}

\begin{figure*}[t]
    \centering{
    \includegraphics[width=1.0\linewidth]{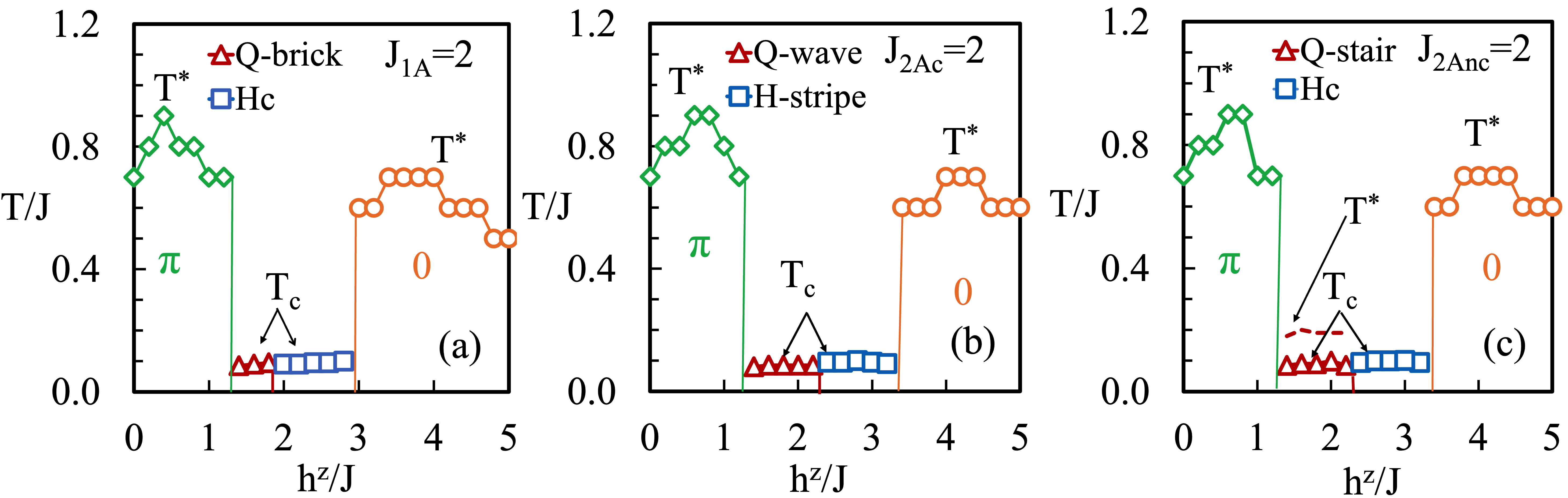}}
    \caption{\label{fig: wide}{Temperature-magnetic field phase diagrams:} (a-c) show the finite temperature phase phases for the three anisotropy cases 1A, 2A$_c$ and 2A$_{nc}$ respectively for the fixed magnitude of anisotropy 2. The $T^*$ indicates crossover temperature scales, while the $T_c$ implies temperature-induced phase phase transition. The phases are labeled in the panels.}
    \label{f-6}
\end{figure*}
\subsection{Temperature evolution}
Next, we discuss the temperature-dependent properties of the vison crystals. In particular, we report the static flux structure factors characterizing the phases as a function of temperature. We particularly inquire if the transition is a continuous or a first-order phase transition for the gapped and the gapless vison crystals. In Fig.~\ref{f-5} (a) and (b), we show the $S(\mathbf{q})$ as a function of temperature for the Q-brick vison crystal for a gapped and gapless parameter point. We find an abrupt transition from the low-temperature ordered gapped Q-brick to a disordered phase, indicating a first-order phase transition for the gapped vison crystal. In sharp contrast, the gapless Q-brick phase appears to be a continuous transition. Similarly, the $S(\mathbf{q})$ peaks for the Q-wave show a first-order (second-order)  transition for the gapless (gapped) phase in Fig.~\ref{f-5} (c) and (d). The gapped phases for the H-stripe and the Q-stairs phases also exhibit a first-order transition.   Interestingly, we observe that for the Q-stair phase in (f) there is an intermediate temperature regime where only the $(\pi/2,\pi/2)$ is suppressed to zero while the ($\pi,\pi$) and the $(0,0)$ phase go to zero at a higher temperature. 

Next, we construct the temperature-magnetic field phase diagrams for the three anisotropy cases as depicted in Fig.~\ref{f-6}. We show these at an anisotropy magnitude of 2 in all three cases. At this anisotropy value, all phases are gapless except the Q-wave phase for 2A$_c$. Consequently, all phase transitions are first order, with only the  Q-wave phase in (b) showing a continuous transition. Since the 0 and the $\pi$ flux configurations do not break translation symmetry, they do not exhibit phase transitions. Instead they show a crossover behavior with a characteristic temperature scale, $T^*$ \cite{kee-1d-mc, burnell-kitaev-mc, motome-3d-mc}. We estimate the crossover temperatures from $S(0)$. The transition to the ordered phase is characterized by reduced discrete translation symmetry. The corresponding $T_c$ values are about an order of magnitude smaller than the crossover temperatures. However, for 2$A_{nc}$, the Q-stair phase has a crossover scale (dashed line in panel (c)) where only $q=(\pi/2,\pi/2)$ and $q=(0,0)$ peaks become finite, while the fully ordered phase occurs below $T_c$ where in addition the $q=(\pi,\pi)$ phase becomes finite. We observe that the $T^*$ scales for the zero and $\pi$ phases show maxima at $h^z$ values in the middle of the respective phases. However, the $T_c$ scales are largely field independent, within our numerical accuracy. 

\section{Conclusion \& Outlook}
We employed a combination of unbiased Monte Carlo simulations and variational analysis to uncover novel gapped and gapless vison crystals for a spin-orbital model with Kitaev-type interactions on a square lattice in the presence of magnetic field. We discovered a new type of vison crystals, quarter phases, wherein a new 16 site periodicity emerge. Our results show gapped vison crystals exhibit a magnetization plateau, while the magnetization grows monotonically with the field for the gapless phases. We have tracked the temperature evolution of the structure factors for the novel phases and uncovered that the gapped phases show an abrupt phase transition in the crystal phase with reducing temperature. In contrast, the gapless vison crystals show a continuous transition from the high-temperature disordered phase to vison crystals. 
The magnetization plateau can be observed easily in standard magnetization measurements, in the presence of a magnetic field, and based on our results, can indicate gapped and gapless phases. The gapped and gapless flux crystals can also show a clear signature in specific heat. Specific heat $C_V\sim e^{-\Delta/K_BT}$  ($\Delta$, be smaller of the gap in the Majorana and the energy gap to vison excitation) for the gapped $C_v$ is expected to scale as $T^2$ and $T$, respectively. Further, neutron scattering experiments can capture the spontaneous breaking of translation symmetry in the various vison crystals. Finally, resonant inelastic spectroscopy can be used to capture signatures of Majorana excitations. 

\section{Acknowledgements}
OE acknowledge support
from NSF Award No. DMR 2234352.

\bibliography{bibliography.bib}

\begin{thebibliography}{29}%
\makeatletter
\providecommand \@ifxundefined [1]{%
 \@ifx{#1\undefined}
}%
\providecommand \@ifnum [1]{%
 \ifnum #1\expandafter \@firstoftwo
 \else \expandafter \@secondoftwo
 \fi
}%
\providecommand \@ifx [1]{%
 \ifx #1\expandafter \@firstoftwo
 \else \expandafter \@secondoftwo
 \fi
}%
\providecommand \natexlab [1]{#1}%
\providecommand \enquote  [1]{``#1''}%
\providecommand \bibnamefont  [1]{#1}%
\providecommand \bibfnamefont [1]{#1}%
\providecommand \citenamefont [1]{#1}%
\providecommand \href@noop [0]{\@secondoftwo}%
\providecommand \href [0]{\begingroup \@sanitize@url \@href}%
\providecommand \@href[1]{\@@startlink{#1}\@@href}%
\providecommand \@@href[1]{\endgroup#1\@@endlink}%
\providecommand \@sanitize@url [0]{\catcode `\\12\catcode `\$12\catcode
  `\&12\catcode `\#12\catcode `\^12\catcode `\_12\catcode `\%12\relax}%
\providecommand \@@startlink[1]{}%
\providecommand \@@endlink[0]{}%
\providecommand \url  [0]{\begingroup\@sanitize@url \@url }%
\providecommand \@url [1]{\endgroup\@href {#1}{\urlprefix }}%
\providecommand \urlprefix  [0]{URL }%
\providecommand \Eprint [0]{\href }%
\providecommand \doibase [0]{http://dx.doi.org/}%
\providecommand \selectlanguage [0]{\@gobble}%
\providecommand \bibinfo  [0]{\@secondoftwo}%
\providecommand \bibfield  [0]{\@secondoftwo}%
\providecommand \translation [1]{[#1]}%
\providecommand \BibitemOpen [0]{}%
\providecommand \bibitemStop [0]{}%
\providecommand \bibitemNoStop [0]{.\EOS\space}%
\providecommand \EOS [0]{\spacefactor3000\relax}%
\providecommand \BibitemShut  [1]{\csname bibitem#1\endcsname}%
\let\auto@bib@innerbib\@empty
\bibitem [{\citenamefont {Broholm}\ \emph {et~al.}(2020)\citenamefont
  {Broholm}, \citenamefont {Cava}, \citenamefont {Kivelson}, \citenamefont
  {Nocera}, \citenamefont {Norman},\ and\ \citenamefont
  {Senthil}}]{Broholm_Science2020}%
  \BibitemOpen
  \bibfield  {author} {\bibinfo {author} {\bibfnamefont {C.}~\bibnamefont
  {Broholm}}, \bibinfo {author} {\bibfnamefont {R.~J.}\ \bibnamefont {Cava}},
  \bibinfo {author} {\bibfnamefont {S.~A.}\ \bibnamefont {Kivelson}}, \bibinfo
  {author} {\bibfnamefont {D.~G.}\ \bibnamefont {Nocera}}, \bibinfo {author}
  {\bibfnamefont {M.~R.}\ \bibnamefont {Norman}}, \ and\ \bibinfo {author}
  {\bibfnamefont {T.}~\bibnamefont {Senthil}},\ }\href {\doibase
  10.1126/science.aay0668} {\bibfield  {journal} {\bibinfo  {journal}
  {Science}\ }\textbf {\bibinfo {volume} {367}},\ \bibinfo {pages} {eaay0668}
  (\bibinfo {year} {2020})}\BibitemShut {NoStop}%
\bibitem [{\citenamefont {Balents}(2010)}]{Balents_Nature2010}%
  \BibitemOpen
  \bibfield  {author} {\bibinfo {author} {\bibfnamefont {L.}~\bibnamefont
  {Balents}},\ }\href {\doibase 10.1038/nature08917} {\bibfield  {journal}
  {\bibinfo  {journal} {Nature}\ }\textbf {\bibinfo {volume} {464}},\ \bibinfo
  {pages} {199} (\bibinfo {year} {2010})}\BibitemShut {NoStop}%
\bibitem [{\citenamefont {Savary}\ and\ \citenamefont
  {Balents}(2016)}]{Savary_RepProgPhys2016}%
  \BibitemOpen
  \bibfield  {author} {\bibinfo {author} {\bibfnamefont {L.}~\bibnamefont
  {Savary}}\ and\ \bibinfo {author} {\bibfnamefont {L.}~\bibnamefont
  {Balents}},\ }\href {\doibase 10.1088/0034-4885/80/1/016502} {\bibfield
  {journal} {\bibinfo  {journal} {Reports on Progress in Physics}\ }\textbf
  {\bibinfo {volume} {80}},\ \bibinfo {pages} {016502} (\bibinfo {year}
  {2016})}\BibitemShut {NoStop}%
\bibitem [{\citenamefont {Moessner}\ and\ \citenamefont
  {Moore}(2021)}]{moessner_moore_2021}%
  \BibitemOpen
  \bibfield  {author} {\bibinfo {author} {\bibfnamefont {R.}~\bibnamefont
  {Moessner}}\ and\ \bibinfo {author} {\bibfnamefont {J.~E.}\ \bibnamefont
  {Moore}},\ }\href {\doibase 10.1017/9781316226308} {\emph {\bibinfo {title}
  {Topological Phases of Matter}}}\ (\bibinfo  {publisher} {Cambridge
  University Press},\ \bibinfo {year} {2021})\BibitemShut {NoStop}%
\bibitem [{\citenamefont {Zhou}\ \emph {et~al.}(2017)\citenamefont {Zhou},
  \citenamefont {Kanoda},\ and\ \citenamefont {Ng}}]{Zhou_RMP2017}%
  \BibitemOpen
  \bibfield  {author} {\bibinfo {author} {\bibfnamefont {Y.}~\bibnamefont
  {Zhou}}, \bibinfo {author} {\bibfnamefont {K.}~\bibnamefont {Kanoda}}, \ and\
  \bibinfo {author} {\bibfnamefont {T.-K.}\ \bibnamefont {Ng}},\ }\href
  {\doibase 10.1103/RevModPhys.89.025003} {\bibfield  {journal} {\bibinfo
  {journal} {Rev. Mod. Phys.}\ }\textbf {\bibinfo {volume} {89}},\ \bibinfo
  {pages} {025003} (\bibinfo {year} {2017})}\BibitemShut {NoStop}%
\bibitem [{\citenamefont {Knolle}\ and\ \citenamefont
  {Moessner}(2019)}]{Knolle_AnnRevCondMatPhys2019}%
  \BibitemOpen
  \bibfield  {author} {\bibinfo {author} {\bibfnamefont {J.}~\bibnamefont
  {Knolle}}\ and\ \bibinfo {author} {\bibfnamefont {R.}~\bibnamefont
  {Moessner}},\ }\href {\doibase 10.1146/annurev-conmatphys-031218-013401}
  {\bibfield  {journal} {\bibinfo  {journal} {Annual Review of Condensed Matter
  Physics}\ }\textbf {\bibinfo {volume} {10}},\ \bibinfo {pages} {451}
  (\bibinfo {year} {2019})}\BibitemShut {NoStop}%
\bibitem [{\citenamefont {Wen}(2017)}]{Wen_RMP2017}%
  \BibitemOpen
  \bibfield  {author} {\bibinfo {author} {\bibfnamefont {X.-G.}\ \bibnamefont
  {Wen}},\ }\href {\doibase 10.1103/RevModPhys.89.041004} {\bibfield  {journal}
  {\bibinfo  {journal} {Rev. Mod. Phys.}\ }\textbf {\bibinfo {volume} {89}},\
  \bibinfo {pages} {041004} (\bibinfo {year} {2017})}\BibitemShut {NoStop}%
\bibitem [{\citenamefont {Kitaev}(2006)}]{kitaev-2006}%
  \BibitemOpen
  \bibfield  {author} {\bibinfo {author} {\bibfnamefont {A.}~\bibnamefont
  {Kitaev}},\ }\href {\doibase https://doi.org/10.1016/j.aop.2005.10.005}
  {\bibfield  {journal} {\bibinfo  {journal} {Annals of Physics}\ }\textbf
  {\bibinfo {volume} {321}},\ \bibinfo {pages} {2} (\bibinfo {year} {2006})},\
  \bibinfo {note} {january Special Issue}\BibitemShut {NoStop}%
\bibitem [{\citenamefont {Cui}\ \emph {et~al.}(2020)\citenamefont {Cui},
  \citenamefont {Ding}, \citenamefont {Han}, \citenamefont {Penington},
  \citenamefont {Ranard}, \citenamefont {Rayhaun},\ and\ \citenamefont
  {Shangnan}}]{Cui_Quantum2020}%
  \BibitemOpen
  \bibfield  {author} {\bibinfo {author} {\bibfnamefont {S.~X.}\ \bibnamefont
  {Cui}}, \bibinfo {author} {\bibfnamefont {D.}~\bibnamefont {Ding}}, \bibinfo
  {author} {\bibfnamefont {X.}~\bibnamefont {Han}}, \bibinfo {author}
  {\bibfnamefont {G.}~\bibnamefont {Penington}}, \bibinfo {author}
  {\bibfnamefont {D.}~\bibnamefont {Ranard}}, \bibinfo {author} {\bibfnamefont
  {B.~C.}\ \bibnamefont {Rayhaun}}, \ and\ \bibinfo {author} {\bibfnamefont
  {Z.}~\bibnamefont {Shangnan}},\ }\href {\doibase 10.22331/q-2020-09-24-331}
  {\bibfield  {journal} {\bibinfo  {journal} {{Quantum}}\ }\textbf {\bibinfo
  {volume} {4}},\ \bibinfo {pages} {331} (\bibinfo {year} {2020})}\BibitemShut
  {NoStop}%
\bibitem [{\citenamefont {Kitagawa}\ \emph {et~al.}(2018)\citenamefont
  {Kitagawa}, \citenamefont {Takayama}, \citenamefont {Matsumoto},
  \citenamefont {Kato}, \citenamefont {Takano}, \citenamefont {Kishimoto},
  \citenamefont {Bette}, \citenamefont {Dinnebier}, \citenamefont {Jackeli},\
  and\ \citenamefont {Takagi}}]{Kitagawa_Nat2018}%
  \BibitemOpen
  \bibfield  {author} {\bibinfo {author} {\bibfnamefont {K.}~\bibnamefont
  {Kitagawa}}, \bibinfo {author} {\bibfnamefont {T.}~\bibnamefont {Takayama}},
  \bibinfo {author} {\bibfnamefont {Y.}~\bibnamefont {Matsumoto}}, \bibinfo
  {author} {\bibfnamefont {A.}~\bibnamefont {Kato}}, \bibinfo {author}
  {\bibfnamefont {R.}~\bibnamefont {Takano}}, \bibinfo {author} {\bibfnamefont
  {Y.}~\bibnamefont {Kishimoto}}, \bibinfo {author} {\bibfnamefont
  {S.}~\bibnamefont {Bette}}, \bibinfo {author} {\bibfnamefont
  {R.}~\bibnamefont {Dinnebier}}, \bibinfo {author} {\bibfnamefont
  {G.}~\bibnamefont {Jackeli}}, \ and\ \bibinfo {author} {\bibfnamefont
  {H.}~\bibnamefont {Takagi}},\ }\href {\doibase 10.1038/nature25482}
  {\bibfield  {journal} {\bibinfo  {journal} {Nature}\ }\textbf {\bibinfo
  {volume} {554}},\ \bibinfo {pages} {341} (\bibinfo {year}
  {2018})}\BibitemShut {NoStop}%
\bibitem [{\citenamefont {Takagi}\ \emph {et~al.}(2019)\citenamefont {Takagi},
  \citenamefont {Takayama}, \citenamefont {Jackeli}, \citenamefont
  {Khaliullin},\ and\ \citenamefont {Nagler}}]{Takagi_NatRevPhys2019}%
  \BibitemOpen
  \bibfield  {author} {\bibinfo {author} {\bibfnamefont {H.}~\bibnamefont
  {Takagi}}, \bibinfo {author} {\bibfnamefont {T.}~\bibnamefont {Takayama}},
  \bibinfo {author} {\bibfnamefont {G.}~\bibnamefont {Jackeli}}, \bibinfo
  {author} {\bibfnamefont {G.}~\bibnamefont {Khaliullin}}, \ and\ \bibinfo
  {author} {\bibfnamefont {S.~E.}\ \bibnamefont {Nagler}},\ }\href {\doibase
  10.1038/s42254-019-0038-2} {\bibfield  {journal} {\bibinfo  {journal} {Nature
  Reviews Physics}\ }\textbf {\bibinfo {volume} {1}},\ \bibinfo {pages} {264}
  (\bibinfo {year} {2019})}\BibitemShut {NoStop}%
\bibitem [{\citenamefont {Lee}\ \emph {et~al.}(2020)\citenamefont {Lee},
  \citenamefont {Utermohlen}, \citenamefont {Weber}, \citenamefont {Hwang},
  \citenamefont {Zhang}, \citenamefont {van Tol}, \citenamefont {Goldberger},
  \citenamefont {Trivedi},\ and\ \citenamefont {Hammel}}]{Lee_PRL2020}%
  \BibitemOpen
  \bibfield  {author} {\bibinfo {author} {\bibfnamefont {I.}~\bibnamefont
  {Lee}}, \bibinfo {author} {\bibfnamefont {F.~G.}\ \bibnamefont {Utermohlen}},
  \bibinfo {author} {\bibfnamefont {D.}~\bibnamefont {Weber}}, \bibinfo
  {author} {\bibfnamefont {K.}~\bibnamefont {Hwang}}, \bibinfo {author}
  {\bibfnamefont {C.}~\bibnamefont {Zhang}}, \bibinfo {author} {\bibfnamefont
  {J.}~\bibnamefont {van Tol}}, \bibinfo {author} {\bibfnamefont {J.~E.}\
  \bibnamefont {Goldberger}}, \bibinfo {author} {\bibfnamefont
  {N.}~\bibnamefont {Trivedi}}, \ and\ \bibinfo {author} {\bibfnamefont
  {P.~C.}\ \bibnamefont {Hammel}},\ }\href {\doibase
  10.1103/PhysRevLett.124.017201} {\bibfield  {journal} {\bibinfo  {journal}
  {Phys. Rev. Lett.}\ }\textbf {\bibinfo {volume} {124}},\ \bibinfo {pages}
  {017201} (\bibinfo {year} {2020})}\BibitemShut {NoStop}%
\bibitem [{\citenamefont {Lieb}(1994)}]{lieb-1994}%
  \BibitemOpen
  \bibfield  {author} {\bibinfo {author} {\bibfnamefont {E.~H.}\ \bibnamefont
  {Lieb}},\ }\href {\doibase 10.1103/PhysRevLett.73.2158} {\bibfield  {journal}
  {\bibinfo  {journal} {Phys. Rev. Lett.}\ }\textbf {\bibinfo {volume} {73}},\
  \bibinfo {pages} {2158} (\bibinfo {year} {1994})}\BibitemShut {NoStop}%
\bibitem [{\citenamefont {Luo}\ \emph {et~al.}(2021)\citenamefont {Luo},
  \citenamefont {Hu},\ and\ \citenamefont {Kee}}]{kee-1d-mc}%
  \BibitemOpen
  \bibfield  {author} {\bibinfo {author} {\bibfnamefont {Q.}~\bibnamefont
  {Luo}}, \bibinfo {author} {\bibfnamefont {S.}~\bibnamefont {Hu}}, \ and\
  \bibinfo {author} {\bibfnamefont {H.-Y.}\ \bibnamefont {Kee}},\ }\href
  {\doibase 10.1103/PhysRevResearch.3.033048} {\bibfield  {journal} {\bibinfo
  {journal} {Phys. Rev. Res.}\ }\textbf {\bibinfo {volume} {3}},\ \bibinfo
  {pages} {033048} (\bibinfo {year} {2021})}\BibitemShut {NoStop}%
\bibitem [{\citenamefont {Feng}\ \emph {et~al.}(2020)\citenamefont {Feng},
  \citenamefont {Perkins},\ and\ \citenamefont {Burnell}}]{burnell-kitaev-mc}%
  \BibitemOpen
  \bibfield  {author} {\bibinfo {author} {\bibfnamefont {K.}~\bibnamefont
  {Feng}}, \bibinfo {author} {\bibfnamefont {N.~B.}\ \bibnamefont {Perkins}}, \
  and\ \bibinfo {author} {\bibfnamefont {F.~J.}\ \bibnamefont {Burnell}},\
  }\href {\doibase 10.1103/PhysRevB.102.224402} {\bibfield  {journal} {\bibinfo
   {journal} {Phys. Rev. B}\ }\textbf {\bibinfo {volume} {102}},\ \bibinfo
  {pages} {224402} (\bibinfo {year} {2020})}\BibitemShut {NoStop}%
\bibitem [{\citenamefont {Nasu}\ \emph {et~al.}(2014)\citenamefont {Nasu},
  \citenamefont {Udagawa},\ and\ \citenamefont {Motome}}]{motome-3d-mc}%
  \BibitemOpen
  \bibfield  {author} {\bibinfo {author} {\bibfnamefont {J.}~\bibnamefont
  {Nasu}}, \bibinfo {author} {\bibfnamefont {M.}~\bibnamefont {Udagawa}}, \
  and\ \bibinfo {author} {\bibfnamefont {Y.}~\bibnamefont {Motome}},\ }\href
  {\doibase 10.1103/PhysRevLett.113.197205} {\bibfield  {journal} {\bibinfo
  {journal} {Phys. Rev. Lett.}\ }\textbf {\bibinfo {volume} {113}},\ \bibinfo
  {pages} {197205} (\bibinfo {year} {2014})}\BibitemShut {NoStop}%
\bibitem [{\citenamefont {Zhang}\ \emph {et~al.}(2019)\citenamefont {Zhang},
  \citenamefont {Wang}, \citenamefont {Hal\'asz},\ and\ \citenamefont
  {Batista}}]{batista-2d-mc}%
  \BibitemOpen
  \bibfield  {author} {\bibinfo {author} {\bibfnamefont {S.-S.}\ \bibnamefont
  {Zhang}}, \bibinfo {author} {\bibfnamefont {Z.}~\bibnamefont {Wang}},
  \bibinfo {author} {\bibfnamefont {G.~B.}\ \bibnamefont {Hal\'asz}}, \ and\
  \bibinfo {author} {\bibfnamefont {C.~D.}\ \bibnamefont {Batista}},\ }\href
  {\doibase 10.1103/PhysRevLett.123.057201} {\bibfield  {journal} {\bibinfo
  {journal} {Phys. Rev. Lett.}\ }\textbf {\bibinfo {volume} {123}},\ \bibinfo
  {pages} {057201} (\bibinfo {year} {2019})}\BibitemShut {NoStop}%
\bibitem [{\citenamefont {Wu}\ \emph {et~al.}(2009)\citenamefont {Wu},
  \citenamefont {Arovas},\ and\ \citenamefont {Hung}}]{Wu_PRB2009}%
  \BibitemOpen
  \bibfield  {author} {\bibinfo {author} {\bibfnamefont {C.}~\bibnamefont
  {Wu}}, \bibinfo {author} {\bibfnamefont {D.}~\bibnamefont {Arovas}}, \ and\
  \bibinfo {author} {\bibfnamefont {H.-H.}\ \bibnamefont {Hung}},\ }\href
  {\doibase 10.1103/PhysRevB.79.134427} {\bibfield  {journal} {\bibinfo
  {journal} {Phys. Rev. B}\ }\textbf {\bibinfo {volume} {79}},\ \bibinfo
  {pages} {134427} (\bibinfo {year} {2009})}\BibitemShut {NoStop}%
\bibitem [{\citenamefont {Chern}(2010)}]{Chern_PRB2010}%
  \BibitemOpen
  \bibfield  {author} {\bibinfo {author} {\bibfnamefont {G.-W.}\ \bibnamefont
  {Chern}},\ }\href {\doibase 10.1103/PhysRevB.81.125134} {\bibfield  {journal}
  {\bibinfo  {journal} {Phys. Rev. B}\ }\textbf {\bibinfo {volume} {81}},\
  \bibinfo {pages} {125134} (\bibinfo {year} {2010})}\BibitemShut {NoStop}%
\bibitem [{\citenamefont {Yao}\ and\ \citenamefont
  {Lee}(2011)}]{yao-lee-flux-lattice}%
  \BibitemOpen
  \bibfield  {author} {\bibinfo {author} {\bibfnamefont {H.}~\bibnamefont
  {Yao}}\ and\ \bibinfo {author} {\bibfnamefont {D.-H.}\ \bibnamefont {Lee}},\
  }\href {\doibase 10.1103/PhysRevLett.107.087205} {\bibfield  {journal}
  {\bibinfo  {journal} {Phys. Rev. Lett.}\ }\textbf {\bibinfo {volume} {107}},\
  \bibinfo {pages} {087205} (\bibinfo {year} {2011})}\BibitemShut {NoStop}%
\bibitem [{\citenamefont {Nakai}\ \emph {et~al.}(2012)\citenamefont {Nakai},
  \citenamefont {Ryu},\ and\ \citenamefont {Furusaki}}]{Nakai_PRB2012}%
  \BibitemOpen
  \bibfield  {author} {\bibinfo {author} {\bibfnamefont {R.}~\bibnamefont
  {Nakai}}, \bibinfo {author} {\bibfnamefont {S.}~\bibnamefont {Ryu}}, \ and\
  \bibinfo {author} {\bibfnamefont {A.}~\bibnamefont {Furusaki}},\ }\href
  {\doibase 10.1103/PhysRevB.85.155119} {\bibfield  {journal} {\bibinfo
  {journal} {Phys. Rev. B}\ }\textbf {\bibinfo {volume} {85}},\ \bibinfo
  {pages} {155119} (\bibinfo {year} {2012})}\BibitemShut {NoStop}%
\bibitem [{\citenamefont {Vijayvargia}\ \emph {et~al.}(2023)\citenamefont
  {Vijayvargia}, \citenamefont {Nica}, \citenamefont {Moessner}, \citenamefont
  {Lu},\ and\ \citenamefont {Erten}}]{Vijayvargia_PRR2023}%
  \BibitemOpen
  \bibfield  {author} {\bibinfo {author} {\bibfnamefont {A.}~\bibnamefont
  {Vijayvargia}}, \bibinfo {author} {\bibfnamefont {E.~M.}\ \bibnamefont
  {Nica}}, \bibinfo {author} {\bibfnamefont {R.}~\bibnamefont {Moessner}},
  \bibinfo {author} {\bibfnamefont {Y.-M.}\ \bibnamefont {Lu}}, \ and\ \bibinfo
  {author} {\bibfnamefont {O.}~\bibnamefont {Erten}},\ }\href {\doibase
  10.1103/PhysRevResearch.5.L022062} {\bibfield  {journal} {\bibinfo  {journal}
  {Phys. Rev. Res.}\ }\textbf {\bibinfo {volume} {5}},\ \bibinfo {pages}
  {L022062} (\bibinfo {year} {2023})}\BibitemShut {NoStop}%
\bibitem [{\citenamefont {Nica}\ \emph {et~al.}(2023)\citenamefont {Nica},
  \citenamefont {Akram}, \citenamefont {Vijayvargia}, \citenamefont
  {Moessner},\ and\ \citenamefont {Erten}}]{onur-moire}%
  \BibitemOpen
  \bibfield  {author} {\bibinfo {author} {\bibfnamefont {E.~M.}\ \bibnamefont
  {Nica}}, \bibinfo {author} {\bibfnamefont {M.}~\bibnamefont {Akram}},
  \bibinfo {author} {\bibfnamefont {A.}~\bibnamefont {Vijayvargia}}, \bibinfo
  {author} {\bibfnamefont {R.}~\bibnamefont {Moessner}}, \ and\ \bibinfo
  {author} {\bibfnamefont {O.}~\bibnamefont {Erten}},\ }\href {\doibase
  10.1038/s41535-023-00541-2} {\bibfield  {journal} {\bibinfo  {journal} {npj
  Quantum Materials}\ }\textbf {\bibinfo {volume} {8}},\ \bibinfo {pages} {9}
  (\bibinfo {year} {2023})}\BibitemShut {NoStop}%
\bibitem [{\citenamefont {Chulliparambil}\ \emph {et~al.}(2021)\citenamefont
  {Chulliparambil}, \citenamefont {Janssen}, \citenamefont {Vojta},
  \citenamefont {Tu},\ and\ \citenamefont {Seifert}}]{vojta-flux-lattice-1}%
  \BibitemOpen
  \bibfield  {author} {\bibinfo {author} {\bibfnamefont {S.}~\bibnamefont
  {Chulliparambil}}, \bibinfo {author} {\bibfnamefont {L.}~\bibnamefont
  {Janssen}}, \bibinfo {author} {\bibfnamefont {M.}~\bibnamefont {Vojta}},
  \bibinfo {author} {\bibfnamefont {H.-H.}\ \bibnamefont {Tu}}, \ and\ \bibinfo
  {author} {\bibfnamefont {U.~F.~P.}\ \bibnamefont {Seifert}},\ }\href
  {\doibase 10.1103/PhysRevB.103.075144} {\bibfield  {journal} {\bibinfo
  {journal} {Phys. Rev. B}\ }\textbf {\bibinfo {volume} {103}},\ \bibinfo
  {pages} {075144} (\bibinfo {year} {2021})}\BibitemShut {NoStop}%
\bibitem [{\citenamefont {Chulliparambil}\ \emph {et~al.}(2020)\citenamefont
  {Chulliparambil}, \citenamefont {Seifert}, \citenamefont {Vojta},
  \citenamefont {Janssen},\ and\ \citenamefont {Tu}}]{vojta-flux-lattice-0}%
  \BibitemOpen
  \bibfield  {author} {\bibinfo {author} {\bibfnamefont {S.}~\bibnamefont
  {Chulliparambil}}, \bibinfo {author} {\bibfnamefont {U.~F.~P.}\ \bibnamefont
  {Seifert}}, \bibinfo {author} {\bibfnamefont {M.}~\bibnamefont {Vojta}},
  \bibinfo {author} {\bibfnamefont {L.}~\bibnamefont {Janssen}}, \ and\
  \bibinfo {author} {\bibfnamefont {H.-H.}\ \bibnamefont {Tu}},\ }\href
  {\doibase 10.1103/PhysRevB.102.201111} {\bibfield  {journal} {\bibinfo
  {journal} {Phys. Rev. B}\ }\textbf {\bibinfo {volume} {102}},\ \bibinfo
  {pages} {201111} (\bibinfo {year} {2020})}\BibitemShut {NoStop}%
\bibitem [{\citenamefont {Keskiner}\ \emph {et~al.}(2023)\citenamefont
  {Keskiner}, \citenamefont {Erten},\ and\ \citenamefont
  {Oktel}}]{onur-quasicrystal}%
  \BibitemOpen
  \bibfield  {author} {\bibinfo {author} {\bibfnamefont {M.~A.}\ \bibnamefont
  {Keskiner}}, \bibinfo {author} {\bibfnamefont {O.}~\bibnamefont {Erten}}, \
  and\ \bibinfo {author} {\bibfnamefont {M.~O.}\ \bibnamefont {Oktel}},\ }\href
  {\doibase 10.1103/PhysRevB.108.104208} {\enquote {\bibinfo {title}
  {Kitaev-type spin liquid on a quasicrystal},}\ } (\bibinfo {year}
  {2023})\BibitemShut {NoStop}%
\bibitem [{\citenamefont {Akram}\ \emph {et~al.}(2023)\citenamefont {Akram},
  \citenamefont {Nica}, \citenamefont {Lu},\ and\ \citenamefont
  {Erten}}]{Akram_PRB2023}%
  \BibitemOpen
  \bibfield  {author} {\bibinfo {author} {\bibfnamefont {M.}~\bibnamefont
  {Akram}}, \bibinfo {author} {\bibfnamefont {E.~M.}\ \bibnamefont {Nica}},
  \bibinfo {author} {\bibfnamefont {Y.-M.}\ \bibnamefont {Lu}}, \ and\ \bibinfo
  {author} {\bibfnamefont {O.}~\bibnamefont {Erten}},\ }\href {\doibase
  10.1103/PhysRevB.108.224427} {\bibfield  {journal} {\bibinfo  {journal}
  {Phys. Rev. B}\ }\textbf {\bibinfo {volume} {108}},\ \bibinfo {pages}
  {224427} (\bibinfo {year} {2023})}\BibitemShut {NoStop}%
\bibitem [{\citenamefont {Erten}\ \emph {et~al.}(2011)\citenamefont {Erten},
  \citenamefont {Meetei}, \citenamefont {Mukherjee}, \citenamefont {Randeria},
  \citenamefont {Trivedi},\ and\ \citenamefont {Woodward}}]{Erten_PRL2011}%
  \BibitemOpen
  \bibfield  {author} {\bibinfo {author} {\bibfnamefont {O.}~\bibnamefont
  {Erten}}, \bibinfo {author} {\bibfnamefont {O.~N.}\ \bibnamefont {Meetei}},
  \bibinfo {author} {\bibfnamefont {A.}~\bibnamefont {Mukherjee}}, \bibinfo
  {author} {\bibfnamefont {M.}~\bibnamefont {Randeria}}, \bibinfo {author}
  {\bibfnamefont {N.}~\bibnamefont {Trivedi}}, \ and\ \bibinfo {author}
  {\bibfnamefont {P.}~\bibnamefont {Woodward}},\ }\href {\doibase
  10.1103/PhysRevLett.107.257201} {\bibfield  {journal} {\bibinfo  {journal}
  {Phys. Rev. Lett.}\ }\textbf {\bibinfo {volume} {107}},\ \bibinfo {pages}
  {257201} (\bibinfo {year} {2011})}\BibitemShut {NoStop}%
\bibitem [{\citenamefont {Nasu}\ \emph {et~al.}(2015)\citenamefont {Nasu},
  \citenamefont {Udagawa},\ and\ \citenamefont {Motome}}]{Nasu_PRB2015}%
  \BibitemOpen
  \bibfield  {author} {\bibinfo {author} {\bibfnamefont {J.}~\bibnamefont
  {Nasu}}, \bibinfo {author} {\bibfnamefont {M.}~\bibnamefont {Udagawa}}, \
  and\ \bibinfo {author} {\bibfnamefont {Y.}~\bibnamefont {Motome}},\ }\href
  {\doibase 10.1103/PhysRevB.92.115122} {\bibfield  {journal} {\bibinfo
  {journal} {Phys. Rev. B}\ }\textbf {\bibinfo {volume} {92}},\ \bibinfo
  {pages} {115122} (\bibinfo {year} {2015})}\BibitemShut {NoStop}%
\end{thebibliography}%

\end{document}